\documentclass[twocolumn,showpacs,preprintnumbers,amsmath,amssymb]{revtex4}

\usepackage{graphicx}
\usepackage{dcolumn}
\usepackage{bm}

\newcommand{\nl}{\nonumber \\ }

\newcommand{\A}{{\cal A}}

\newcommand{\C}{{\cal C}}

\newcommand{\be}{\begin{equation}}  
\newcommand{\ee}{\end{equation}}  
\newcommand{\bear}{\begin{eqnarray}}  
\newcommand{\eear}{\end{eqnarray}}  
\newcommand{\ba}{\begin{array}}  
\newcommand{\ea}{\end{array}}

\newskip\humongous \humongous=0pt plus 1000pt minus 1000pt

\newif\ifdtup

\def\oldreffmt#1{\rlap{[#1]} \hbox to 2\parindent{}}

\def\figfmt#1{\rlap{Figure {#1}} \hbox to 1in{}}  
  
\def\ie{\hbox{\it i.e.}{}}      \def\etc{\hbox{\it etc.}{}}  
\def\eg{\hbox{\it e.g.}{}}

\def\Tr{\mathop{\rm Tr}}

\def\slash#1{#1\!\!\!/\!\,\,}  
\def\beq{\begin{equation}}  
\def\eeq{\end{equation}}  
\def\bea{\begin{eqnarray}}  
\def\eea{\end{eqnarray}}  
\def\half{\frac{1}{2}}  
  
\def\bq{\begin{quote}}  
\def\eq{\end{quote}}

\def\half{\frac{1}{2}}       
\def \lta {\mathrel{\vcenter  
     {\hbox{$<$}\nointerlineskip\hbox{$\sim$}}}}  
   
\relax  

\newdimen\tdim  
\tdim=\unitlength  
\def\bar{\overline}

\begin{document}  

\preprint{EFI Preprint 07-27}
\preprint{FERMILAB-PUB-07-628-T}
%
%
\title{
Standard Model Gauging of the WZW Term: \\ 
Anomalies, Global Currents and pseudo-Chern-Simons
Interactions}

\author{
Jeffrey A. Harvey$^{(a)}$, 
Christopher T. Hill$^{(b)}$, 
Richard J. Hill$^{(b)}$
}

\thanks{Electronic address: 
harvey@theory.uchicago.edu, hill@fnal.gov, rjh@fnal.gov}

\affiliation{\vspace{0.2in}
$^{(a)}$Enrico Fermi Institute and Department of Physics \\
The University of Chicago, Chicago, Illinois, 60637, USA\\ 
$^{(b)}$Fermi National Accelerator Laboratory\\
\it P.O. Box 500, Batavia, Illinois 60510, USA
}

\date{\today}

\begin{abstract} 
The standard model $SU(2)_L\times U(1)_Y$ gauging of the
Wess-Zumino-Witten term requires a modified counterterm when
background fields, needed to generate the full set of currents,
are introduced.  The modified counterterm plays an essential role in
properly defining covariant global currents and their anomalies.  
For example, it is required in order to correctly derive 
the gauge invariant baryon number current and 
its anomalous divergence.  The background fields
can also be promoted 
to a description of the physical spin-1 vector and
axial-vector mesons in QCD and the counterterm leads to 
novel interactions.  These are
(pseudo-) Chern-Simons terms, such as $\epsilon^{\mu\nu\rho\sigma}
\omega_\mu Z_\nu \partial_\rho A_\sigma$ and
$\epsilon^{\mu\nu\rho\sigma} \rho^{\pm}_\mu W^{\mp}_\nu \partial_\rho
A_\sigma$ that mediate new interactions between neutrinos and photons
at finite baryon density.
 \end{abstract}

\maketitle

\section{\label{sec:intro}Introduction}

The low-energy spectrum of QCD contains pseudoscalar mesons
interpreted as the Nambu-Goldstone bosons (NGB's) of spontaneously broken
chiral symmetry.  $U(N_f)_L \times U(N_f)_R$ breaks to the diagonal,
vector subgroup $U(N_f)_V$, with $N_f=2,3$ depending on whether just
the $(u,d)$ quark symmetries, or the $(u,d,s)$ symmetries are included in the
analysis.  A complete low-energy chiral lagrangian describing the
interactions of these meson states contains terms in the following
three classes.

The first class consists of terms related to the familiar kinetic term:
\be 
{\cal L}_K = {f_\pi^2\over 4} {\rm Tr}(D_\mu U^\dagger D^\mu U) +
\dots \,, 
\ee 
where $f_\pi \approx 93\,{\rm MeV}$, and
$U=\exp[(2i/f_\pi) \pi^a T^a]$ is a chiral matrix field transforming
as $U\rightarrow e^{i\epsilon_L}Ue^{-i\epsilon_R}$ under $U(N_f)_L
\times U(N_f)_R$, with $e^{i\epsilon_{L,R}} \in U(N_f)_{L,R}$.  The
kinetic term can be made locally invariant under $U(N_f)_L \times
U(N_f)_R$ transformations by including a complete set of gauge fields
in the covariant derivative, $DU = \partial U - iA_L U + iU A_R$,
with corresponding local gauge transformations for $A_{L,R}$. The
ellipsis refers to an expansion in the number
of derivatives, containing in the next order 
the Gasser-Leutwyler operators.

A second class of terms consists of those associated with symmetry
breaking. This includes the operator $\Tr(M_q U)+h.c.$ where $M_q$ is
the quark mass matrix, and also the operator $\det{U}e^{i\theta} +
h.c.$ 
which reflects the breaking of the axial $U(1)$ by instantons.
We will largely ignore the effects of these first
two classes of operators in what follows.

A third class of operators comprise the Wess-Zumino-Witten (WZW) term,
$\Gamma_{WZW}(U)$~\cite{Wess,Witten}.  This is a topological 
object and it
arises ``holographically'' when the $D=4$ manifold
of spacetime is viewed as the boundary of a $D=5$ manifold \cite{Witten}.
The WZW term is intimately connected to the anomaly structure of
QCD.  When coupled to 
classical background gauge fields $ A_L, A_R
\in U(N_f)_{L,R}$ the variation of $\Gamma_{WZW}(U, A_L, A_R)$
under local $U(N_f)_L \times U(N_f)_R$ is non-zero, and reproduces the
anomalies of the underlying theory of
quarks~\cite{steinberger,BJ,Adler,Bardeen,Wess,Witten,KRS,Manohar:1984uq,Kawai:1984mx}.

Moreover, $\Gamma_{WZW}$ lifts a
spurious parity symmetry in the chiral lagrangian, 
locking the pion 
parity to that of space (it performs a similar task in Little Higgs
theories by breaking spurious $T$-parity~\cite{Hill2}).  It mediates
processes such as $K \bar{K} \rightarrow 3 \pi $ which are
allowed by QCD but would be forbidden in the low-energy chiral
lagrangian theory if we kept only the first two classes of
terms and ignored the WZW term.
The WZW term can also be coupled to physical gauge fields, 
like the photon.
This leads to a correct description of the process $\pi^0 \rightarrow
2\gamma$, that is otherwise forbidden by the extension of the
spurious parity to gauge fields.  
Thus the WZW term generates an essential part of the physics,
and should be placed on the same footing as the other terms in the
chiral lagrangian.

In this paper we are interested in the gauged WZW term with physical
gauge fields coupled to 
the flavor symmetries of the quarks.  In general, to
achieve an anomaly free gauge theory we can either gauge an anomaly
free subgroup of $U(N_f)_L \times U(N_f)_R$, or cancel anomalies
between the chiral lagrangian and a ``lepton sector''.  The former
case arises when gauging only electromagnetism in the QCD chiral
Lagrangian, and is perhaps the most familiar application of a gauged
WZW term.  However, it is the latter situation that arises for the
$SU(2)_L\times U(1)_Y$ electroweak gauge group of the standard model.
This leads to additional issues that need to be
addressed, due to the fact that $SU(2)_L\times
U(1)_Y$ resides in a nondiagonal subgroup of the chiral symmetry group.
We are forced to revisit the counterterm
structure of the Wess-Zumino-Witten term.  Ultimately we are led to a
new counterterm, and in turn, to new physics. This is the focus of the
present paper.

In addition to the fundamental gauge fields of the
standard model, \ie, the $W$, $Z$ and $\gamma$, the theory must 
be consistent if we include background fields that couple to the
currents of the chiral lagrangian. These background fields are
theoretically essential because they allow us to determine the
correct form of the global chiral currents and their anomalies.  

In what follows we will
denote generic fundamental gauge fields by $A$ and background vector
and axial-vector fields by $B$.  The $B$ fields
will be assumed to transform covariantly under the fundamental local
gauge group of the standard model.
Varying the effective action with respect to $B^a$ 
generates the associated current
$J^a$.  Varying the background fields locally, 
$\delta B = d\epsilon + ...$, as
if they were fundamental gauge fields, generates the anomalous
divergence of the associated global current
via the WZW term.
For example, a background field can be introduced
with the quantum numbers of the  $\omega$ meson, 
coupling to the quark baryon current.  This automatically implies
its coupling to the Goldstone-Wilczek skyrmionic baryon number current
\cite{GW,Witten} via the WZW term.

Introducing the $B$ fields leads, however, to the following subtle
issue.  When we have a set of fundamental gauge fields $A$, such
as the standard model $W$, $Z$ and $\gamma$, and we then turn on the
background ($B$) fields, we find that  new anomalies appear in
the gauged ($A$) currents that were previously absent.  We can maintain the
fundamental ($A$) gauge invariance, however, if we can find 
a local counterterm, a functional of $A$ and $B$, 
which cancels these new anomalies.

The logic of this situation is identical to that of QED,
underlying the original
covariant anomaly first computed by Adler~\cite{Adler}. 
If we compute
triangle diagrams for QED with a vector photon $A_\mu$ coupled
to a massless electron as
$A^\mu\bar{\psi}\gamma_\mu\psi$ 
then we do obtain a conserved
vector current, and the divergence of the 
axial current is the {\em consistent} 
axial current anomaly \footnote{Weyl spinor triangle loops 
yield a finite and unambiguous
form of the anomaly that satisfies the Wess-Zumino consistency 
condition, hence known as the consistent anomaly. 
Dirac fermions have ambiguities arising from the left-right
mixing via the mass term. Bardeen obtains the consistent anomaly
for massive Dirac fermions by imposing the left-right symmetry of
the classical Dirac action on the result \cite{Bardeen}. The counterterm
can be viewed as a definition of the fermion loop.
Throughout
this paper we will always assume that the fermion loops explicitly generate 
the consistent anomaly, and that the counterterm is appended to
the action, since this is the appropriate implementation
for the Wess-Zumino-Witten term. 
The counterterm in the action can be seen as arising from
a Chern-Simons term in a $D=5$ theory with chiral delocalization
of fermions on boundary branes,
compactifying the theory to $D=4$ \cite{cth}.}:
\be
\partial^\mu  j^5_\mu = \frac{1}{48\pi^2}
\epsilon_{\mu\nu\rho\sigma}F^{\mu\nu}{F}^{\rho\sigma}.
\ee
However, if we introduce a classical background field $B_\mu$ with
the  coupling $B^\mu j_\mu^5$,
then we find that the vector
current is no longer conserved, but 
develops a mixed anomaly  
$\propto \epsilon_{\mu\nu\rho\sigma} F_A^{\mu\nu}{F}_B^{\rho\sigma}$. 
It is essential that gauge invariance, \ie, vector current
conservation, be maintained for any background field $B$,
and we thus require a counterterm. 
The counterterm takes the form $(1/6\pi^2)\epsilon_{\mu\nu\rho\sigma}
B^\mu A^\nu\partial^\rho A^\sigma$. When added to the action,
it modifies the
definitions of the currents.
For example, the gauged vector current is $\delta S/\delta A_\mu$, 
and the global 
axial current is $\delta S/\delta B_\mu$, where $S$ is the action.  
The currents acquire
corrections from the counterterm.  This 
leads to a conserved modified vector current
for any background field $B$
and the familiar {\em covariant} anomaly for the modified axial current
\footnote{In the case of QED, 
the covariant anomaly is distinguished from
the consistent anomaly only by the factor of $3$ in its coefficient. 
In a Yang-Mills theory the
consistent anomaly is not gauge covariant, while the covariant anomaly is
\cite{Bardeen}. }, 
\be
\partial^\mu j_\mu^5 = 
\frac{1}{16\pi^2}\epsilon_{\mu\nu\rho\sigma}F^{\mu\nu}{F}^{\rho\sigma}.
\ee 

The problem of maintaining gauge invariance of the WZW term 
for arbitrary background fields when the fundamental gauge
fields are vectorlike, \ie,  $A_L = A_R = A$, 
has a well-known solution: the ``Bardeen
counterterm,'' given by: $-\Gamma_{WZW}(1, A +B_L, A+B_R)$.  
This is essentially a generalization of the aforementioned counterterm of QED.
Adding this counterterm to the WZW term, $\Gamma_{WZW}(U, A +B_L,
A+B_R)$, ensures that vector currents are conserved for any
$B_{L,R}$ background fields.  We note that this procedure kills off an
entire class of non-pionic interactions in the bare WZW term, such
as $\epsilon_{\mu\nu\rho\sigma}A^\mu B^\nu_{L,R} \partial^\rho
A^\sigma + ...$,
which we call ``pseudo-Chern-Simons" (pCS) terms~%
\footnote{Interactions of this form are often referred 
to as ``Chern-Simons'' terms, though technically
this is a misnomer. Chern-Simons terms occur only in odd dimensions and
generate anomalies on boundaries. 
In even $D$ we refer to terms like 
$V^{(1)} V^{(2)} dV^{(3)}$
as {\em pseudo-Chern-Simons} terms (or pCS terms) which generate
anomalies in the bulk.
}.  The special case of vectorlike gauging may thus 
lead to the intuition that
pCS terms are somehow unphysical and don't appear in the full theory.
However, this would be a false impression.
 
When gauging nondiagonal 
(non-vectorlike) subgroups such as $SU(2)_L\times
U(1)_Y$, the Bardeen counterterm does not render the theory gauge ($A$) anomaly
free.  We will show, however, that there always exists a 
new local
counterterm that does maintain gauge invariance for the
gauging of any
subgroup in the general background of spin-1 classical ($B$) fields.  We
give the explicit solution for the new counterterm in the general case
and apply it in specific cases.
 
Once the new counterterm is incorporated into the WZW term two
important things happen.  First, the  global currents, as
generated by local variations of the appropriate background
fields, become proper covariant objects.  The
anomalies of these currents are the covariant anomalies of the theory.   
The global baryon current and its anomaly provides an important example of this
phenomenon.  The current is 
modified from the Goldstone-Wilczek form, 
in the presence of gauge fields, and becomes a gauge 
invariant operator.
Its anomaly, arising from a local gauge transformation of the WZW term in the
background $\omega$ field, $\delta \omega_\mu = \partial_\mu \epsilon$,
yields the correct covariant baryon current anomaly.
We note that there are also corrections involving the
background fields themselves, \eg, including a 
term $\epsilon^{\mu\nu\rho\sigma}
F^a_{W\mu\nu} D_\rho \rho^a_\sigma$ where $\rho^a_\mu$ is the
background field with quantum numbers of the $\rho$-meson 
and $F_{W\mu\nu}^a$ is the $SU(2)_L$ field strength. 

Second, there are now uncancelled pCS term interactions involving the
fundamental gauge fields and the spin-1 background fields
that do not involve the pions.  These pCS terms contain 
observable new physics.  

Indeed, the classical background ($B$) fields can be
promoted to describe 
the physical vector meson fields of QCD, 
\ie, the $\rho$, $\omega$, $a_1$, $f_1$ and so
on.  It is important to realize the distinction between classical
background fields and physical spin-1 mesons:
the former would describe ``pointlike
particles,'' present on all scales of
the theory, while the physical
spin-1 mesons have form-factors and decouple from the high energy
quark loops.  They are only part of an effective low energy
theory. However, at low energies the physical spin-1 
mesons can be viewed as coupling to the global currents 
and they thus behave like the $B$
fields.   Anomalous physical processes involving them
can be described by the WZW term.  The pCS terms are a new part
of the physics in the WZW term, involving 
exclusively the spin-1 mesons and gauge fields.

There are many formal issues that must be faced in the description of the
vector mesons as propagating physical particles, and treating them  as
gauge fields~\cite{KRS,Harada} incurs subtleties.  
However, phenomenologically successful treatments of processes that 
involve the spin-1 mesons and probe the
anomalies encoded into the WZW term, such as $\omega,\rho \to \pi
\gamma$, $\omega\rightarrow 3\pi$, \etc., do flourish in the
literature.  We will postpone the detailed discussion of these issues
to a subsequent paper~\cite{HHH1}.

Notably, from the new counterterm we obtain an
interaction of the form $\epsilon^{\mu\nu\rho\sigma}\omega_\mu Z_\nu
F_{\rho\sigma}$ where $\omega$ is the omega meson background field,
$Z$ the $Z$-boson in unitary gauge, and $F_{\mu\nu}$ the photon field
strength~\cite{HHH2}.  This interaction survives as an
essential consequence of the non-diagonal standard model gauge structure
and the new counterterm.

The outline of this paper is as follows.  In Section II we construct a
schematic version of the standard model, \ie, a ``toy'' model, in
which the WZW term is nontrivial, but much simpler than in the
standard model.  This model consists of a single color and flavor of
quark, and a single lepton. We gauge the $U(1)_L\times U(1)_R$ quark
and lepton flavor symmetries by introducing a ``$Z$'' associated with
$U(1)_L$ and a photon ``$A$'' associated with $U(1)_{V}$. The gauge
anomalies cancel between the quark and the lepton sectors, as in the
standard model.  We then integrate out the quark with a large chirally
invariant ``constituent'' mass term, $m_q\bar{q}_Lq_Re^{i\phi/f}$,
containing a ``pion'' $\phi$.  This generates the WZW term involving
$\phi$, $Z$ and $A$, which is easy to derive.

We then introduce the ``$\omega$'' vector meson as a background field
coupled to the baryon current.  We show that new anomalies arise in
the gauged currents and then construct the counterterm that cancels
these anomalies.  We discover that pCS terms such as
$\epsilon_{\mu\nu\rho\sigma}\omega^\mu Z^\nu F^{\rho\sigma}$ remain in
the physical WZW term \cite{HHH2}.  Variation of the $\omega$ field
generates the global baryon current and associated covariant anomaly.

In Section III we consider the general problem of a chiral lagrangian
for a theory in which the chiral flavor symmetry $G$ is spontaneously
broken to a subgroup $H$ while at the same time we gauge a subgroup
$G'\subset G$. We show how to construct the counterterm that maintains
the $G'$ gauge anomaly structure in the presence of background
spin-$1$ fields.  For a diagonal gauge group $G'\subset H$, this
reduces to the Bardeen counterterm (modulo gauge invariant operators).

In Section~\ref{sec:cov} we show that this counterterm plays a crucial
role in the derivation of global symmetry currents and their
anomalies.  The anomalous baryon current provides an important
application of this formalism.  At the chiral lagrangian level we
obtain a gauge invariant baryon current from the WZW term with
the new counterterm.  We examine the
global symmetries that are neutral under the gauged symmetries (\ie,
for which there is no explicit symmetry breaking by gauging), and find
the general form of the global anomalies for arbitrary background
fields.
 
In Section~\ref{sec:sm} we apply these ideas to derive the WZW term
(including counterterms) for the $SU(2)_L\times U(1)_Y$ gauging of the
$U(2)_L\times U(2)_R$ chiral symmetry of QCD in a background of the
spin-1 vector mesons $\rho$, $\omega$, $a_1$, and $f_1$.  Physical
applications of these ideas are mentioned, but the details are
postponed to a subsequent paper \cite{HHH1}.  For example, anomaly
mediated neutrino photon interactions arise from the
$\epsilon_{\mu\nu\rho\sigma}\omega^\mu Z^\nu F^{\rho\sigma}$ pCS
interaction, and provide a possible explanation for excess 
events seen in the 
MiniBooNE experiment \cite{MiniBooNE,HHH2,HHH1}.  In
Section~\ref{sec:conc} we conclude and outline some further
implications of these ideas.

The $SU(2)_L\times U(1)_Y$ gauging of the QCD WZW term in general
backgrounds has not, to our knowledge, been previously developed.
Pseudo-Chern Simons terms with arbitrary coefficients have previously
been appended to the effective Lagrangian in an {\it ad hoc}
manner~\cite{Kaiser, Klingl, Truhlik}, with various phenomenological
constraints on the coefficients.  The advantage of our approach is
that we predict the coefficients of such interactions in terms of the
strong coupling constants of the QCD vector and axial-vector mesons.
Our observations about pCS terms and global anomalies 
apply to general chiral lagrangian models, and are new.

\section{\label{sec:toy}Schematic Standard Model with pCS interactions}
 
\subsection{The WZW term}

We now construct a schematic model that exhibits in a simple way the
necessity of adding new counterterms to the WZW term, and the
existence of pCS terms.  This model involves only abelian gauge
groups, but is constructed in close analogy to QCD and the
$SU(2)_L\times U(1)_Y$ electroweak sector of the standard model.  It
will form the basis for the general discussion of the $SU(2)_L\times
U(1)_Y$ gauging of the $U(2)_L\times U(2)_R$ chiral lagrangian of
pions.

We consider a theory with a single ($N_c=1$) ``quark'' $q$ and a
single ``lepton'' $\ell$. We introduce $U(1)_L$ and $U(1)_R$
fundamental gauge fields $A_L$ and $A_R$ into the quark action: 
\beq
\label{q}
S_q = \int d^4x\;\;\bar{q}_L(i\slash{\partial} + \slash{A}_L)q_L 
+ \bar{q}_R(i\slash{\partial} + \slash{A}_R)q_R \,.
\eeq
$S_q$ has gauge currents,
\bea
J^{(B)_{L}}_{\mu} &=& \frac{\delta S_q}{\delta A_{L}^{\mu}} = \bar{q}_L\gamma_\mu q_L \,,
\nonumber \\
J^{(B)_{R}}_{\mu} &=& \frac{\delta S_q}{\delta A_{R}^{\mu}} = \bar{q}_R\gamma_\mu q_R \,,
\eea
that are anomalous:
\bea
\label{origanom}
\partial^\mu J^{(B)_L}_{\mu} 
&=& 
-\frac{1}{24\pi^2}
\epsilon_{\mu\nu\rho\sigma} 
\partial^\mu A_L^\nu \partial^\rho A_L^\sigma \,, 
\nonumber \\
\partial^\mu J^{(B)_R}_{\mu} 
&=& 
\frac{1}{24\pi^2}
\epsilon_{\mu\nu\rho\sigma} 
\partial^\mu A_R^\nu \partial^\rho A_R^\sigma \,. 
\eea
Here we use the consistent anomalies that arise from
Weyl spinor triangle diagrams.

To cancel these anomalies we gauge the ``lepton'' sector: 
\beq
\label{ell}
S_\ell = \int d^4x\;\;
\bar{\ell}_L(i\slash{\partial} - \slash{A}_L)\ell_L 
+ \bar{\ell}_R(i\slash{\partial} - \slash{A}_R)\ell_R \,.
\eeq 
Note that the relative signs of the quark and lepton couplings
imply that $A_L$ couples to $B-L$ for the left-handed fields, which we
denote by $(B-L)_L$, while $A_R$ couples to $(B-L)_R$.  Taken together
the gauge anomalies cancel between the quark and lepton sectors in the
$B-L$ currents: 
\bea
\label{newanom0}
\partial_\mu(\bar{q}\gamma^\mu q_L - \bar{\ell}\gamma^\mu \ell_L) 
= \partial^\mu
J^{(B-L)_L}_{\mu} = 0 \,,
\nonumber \\
\partial_\mu(\bar{q}\gamma^\mu q_R - \bar{\ell}\gamma^\mu \ell_R) = 
\partial^\mu J^{(B-L)_R}_{\mu} = 0 \,. 
\eea
Anomalies remain in the ungauged $B+L$ currents, imitating the
structure of the standard model.

We are interested in an analogy
to hadronic physics and the chiral lagrangian of QCD. Thus,
we want to spontaneously
break the $U(1)_L\times U(1)_R$ of the quark sector
to $U(1)_V$.  We can do so by introducing 
a constituent quark mass term containing an NGB denoted by $\phi$:
\beq
\label{phiphi}
m_q e^{i\phi/f} \bar{q}_L q_R  + h.c. \,.
\eeq 
Here $\phi/f$ is the analog of $\pi/f_\pi$ in QCD.

Technically, in this model the NGB, $\phi$, would be eaten by the
linear combination $Z = A_L - A_R$, which then becomes massive.  This
is the analog of symmetry breaking in technicolor theories.
Alternatively, we can imagine an additional Higgs scalar field that
gives the $Z$ its mass by developing a VEV $v$. $Z$ then acquires a
longitudinal component, $\chi$, from the phase of the Higgs,
$Z\rightarrow Z - \partial\chi/v$. In this case a dynamical NGB
remains in the low-energy spectrum which is a linear combination of
$\chi$ and $\phi$ (mainly the $\phi$ field in the $v \gg f$ limit). We
will assume the Higgs mechanism is present, allowing the $Z$ to
acquire mass, but we need not explicitly write the $\chi$ lagrangian.

In what follows we will use the abbreviated notation of
differential forms, so that for example 
$\int d^4x\; 
\epsilon_{\mu\nu\rho\sigma}A^\mu B^\nu \partial^\rho C^\sigma = \int
ABdC$.

Under $U(1)_L\times U(1)_R$ gauge transformations
we have:
\bea
\label{gtq}
q_L & \rightarrow & e^{i\epsilon_L} q_L  \,, \qquad 
\ell_L \rightarrow  e^{-i\epsilon_L}\ell_L \,, \qquad
\delta A_L  = d\epsilon_L  \,, 
\nl
q_R & \rightarrow & e^{i\epsilon_R} q_R \,,  \qquad
\ell_R \rightarrow e^{-i\epsilon_R}\ell_R \,, \qquad
\delta A_R  =  d\epsilon_R \,, 
\nl
\delta \phi &=& f(\epsilon_L-\epsilon_R) \,. 
\eea
The gauge transformations acting purely on the
quark sector are anomalous and induce a
shift in the quark effective action, $\delta S
=\int (\partial_\mu \epsilon) J^\mu = -\int \epsilon\, \partial_\mu J^\mu$:
\bea
\label{anom}
\delta S_q =  \frac{1}{24\pi^2} \int \epsilon_L dA_L dA_L
- \epsilon_R dA_R dA_R \,.
\eea 
This is, of course, cancelled by the anomalous shift in the
lepton effective action: 
\beq 
\delta S_\ell = -\frac{1}{24\pi^2}
\int \epsilon_L dA_L dA_L
- \epsilon_R dA_R dA_R \,.
\eeq 
Hence, overall we have a non-anomalous gauge symmetry and
conserved gauged currents as stated in Eq.(\ref{newanom0}).

We now consider a large $m_q$ limit and integrate out the quarks
(which imitates the effect of confinement;
related examples have been discussed in \cite{dhokerI,dhokerII}). 
We are left with an effective action:
\beq 
\Gamma_{WZW}(U, A_L, A_R) + ... \,, 
\eeq 
where $\Gamma_{WZW}$ is the Wess-Zumino-Witten term and
the ellipsis refers
to non-topological terms, such as renormalized $\phi$ kinetic terms.
$\Gamma_{WZW}$ is a functional
of $U =  e^{i\phi/f} $ and the gauge fields, 
$A_L$ and $A_R$.  $\Gamma_{WZW}$ generates
the same anomalies as the quark action in Eq.(\ref{anom})
under the gauge transformations of Eq.(\ref{gtq}).

It is easy to construct the WZW term,  
by arranging a set of operators that generate
the independent $L$ and $R$ consistent anomalies.
We readily obtain:
\begin{multline}
\label{WZW1}
\Gamma_{WZW} = \frac{1}{24\pi^2} \int \;
\bigg[ A_L A_R dA_L + A_L A_R dA_R  
\\
+ \frac{\phi}{f}(dA_LdA_L + dA_R dA_R + dA_L dA_R )\bigg]
\,. 
\end{multline}
It can easily be checked that, under the gauge transformations 
(\ref{gtq}), we have:
\be
\delta \Gamma_{WZW} = \delta S_q \,,
\ee
with $\delta S_q$ from Eq.(\ref{anom}).
 
Note that 
Eq.(\ref{WZW1}) can be obtained from the expression for the 
$U(N)\times U(N)/U(N)$ WZW term discussed in Ref.~\cite{KRS}, 
by taking $N=1$. 
It can also be straightforwardly derived 
``holographically'' from the Chern-Simons
term and Dirac determinant of a compactified $D=5$ $U(1)$ gauge theory
in which $\phi \sim A_5$,  as in Ref.~\cite{cth}.

\subsection{Introduction of the $\omega$}

We now introduce a classical
background field coupled to the ``baryon number'' 
in the quark sector.  We denote this field by  $\omega$ in 
analogy to the $\omega$ meson
of QCD which couples to the baryon current.  The WZW action 
becomes:
\beq
\label{qq}
\Gamma_{WZW}(U, A_L+\omega, A_R+\omega)  
\eeq
Note that $\omega$ is invariant under $U(1)_L\times U(1)_R$ gauge
transformations. 
In analogy to QCD we view
$\omega$ as part of the strong interactions, 
and do not couple it to the lepton sector. 
With $\omega$ appearing only in the quark sector, we then find  
that the theory now contains anomalies under
local $U(1)_L\times U(1)_R$ gauge transformations.

From Eq.(\ref{anom}) we see that:
\begin{multline}
\label{nnewanom}
\delta (\Gamma_{WZW} + S_\ell )
= 
{1\over 24\pi^2} \int \epsilon_L \left[ 2 dA_L d\omega + (d\omega)^2 \right] 
\\
- \epsilon_R \left[ 2 dA_R d\omega + (d\omega)^2 \right] \,.
\end{multline}
The gauge symmetry, and the internal consistency of the theory,
is apparently spoiled by the inclusion of $\omega$.
If, however, we can find a
local counterterm, a functional of $A_L$ $A_R$ and $\omega$,
to add to the lagrangian that restores $U(1)_L
\times U(1)_R$ gauge invariance in the presence of a background
$\omega$, then the theory can be made consistent,
as in the case of QED, summarized in the Introduction.

The desired counterterm is readily constructed:
\bea
\label{g0full}
\Gamma_c & =& -\frac{1}{24\pi^2}  
\int \; [-2\omega  A_R dA_R - \omega A_R d\omega
\nonumber \\
& & \qquad \qquad
+2\omega  A_L dA_L +  \omega  A_L d\omega ]  \,. 
\eea
This counterterm is a necessary part of the low energy theory
when $\omega$ is introduced into $\Gamma_{WZW}$. 
Adding $\Gamma_c$ to $\Gamma_{WZW}$ we see that 
the new $\omega$ dependent terms in Eq.(\ref{nnewanom}) 
are now cancelled under a gauge transformation. 
If $\omega$ is an arbitrary classical
background field that couples also to quarks at high energies, 
then this counterterm
is required in the high energy action, $S_q$, as well. 

The full WZW term
is now given by the sum of $\Gamma_{WZW}$ and 
the counterterm:
\bea
\label{gfull}
\Gamma_{WZW}^{\rm full} & = & 
\Gamma_{WZW}(\phi, A_L+\omega, A_R+\omega )  + \Gamma_c(A_L,A_R,\omega)
\nonumber \\
& \equiv & 
\Gamma_{WZW}(\phi, A_L, A_R ) + \Gamma_\omega(\phi,A_L,A_R,\omega)  \,, 
\eea
where we've isolated the interactions involving $\omega$ into $\Gamma_\omega$: 
\bea
\label{gfull2}
\Gamma_\omega &  = & 
  \frac{1}{8\pi^2} \int  \;
\bigg[\; \frac{\phi}{f} (dA_L d\omega + dA_R d\omega  + d\omega d\omega)
\nonumber \\
& & - \omega A_L dA_L + \omega A_R dA_R 
 + \omega A_R  dA_L 
\nonumber \\
& & - \omega  A_L dA_R
 +  \omega A_R d\omega   - \omega A_L d\omega   \;\; \bigg] \,. 
\eea
We thus see that pCS terms, such as $\omega A_L dA_L$, 
now appear in the complete effective action.
$\Gamma_{WZW}(\phi, A_L, A_R )$ generates the
original anomalies  $\propto -dA_LdA_L + dA_RdA_R$ 
that are cancelled by the leptons.  
$\Gamma_\omega$  governs interactions of $\phi$, $A_L$ and $A_R$
with $\omega$ 
(these are analogous to anomalous
interactions in QCD such as $\omega\rightarrow \pi^0 \gamma$).

Since $\Gamma_\omega$ generates no new gauge anomalies, it must be itself
a gauge invariant operator.
Moreover, under local shifts in $\omega$, \ie,
$\delta\omega = d\epsilon$ this term
generates the global baryon current anomaly:
\begin{multline}
\label{glo1}
\delta \Gamma_{WZW}^{\rm full}  =  \delta \Gamma_{\omega}  = \\
  \frac{1}{8\pi^2} \int  \; \epsilon
\big[\;dA_L dA_L - dA_R dA_R  + dA_L d\omega 
   -dA_R d\omega    \;\big] \,, 
\end{multline}
or:
\bea
\label{Bcur}
\partial_\mu J^\mu  & =  & 
-\frac{1}{8\pi^2}\epsilon_{\mu\nu\rho\sigma} 
\big[\;\partial^\mu A^\nu_L \partial^\rho A^\sigma_L - 
\partial^\mu A^\nu_R \partial^\rho A^\sigma_R  
 \nonumber \\
& & \qquad + 
   \partial^\mu A^\nu_L \partial^\rho \omega^\sigma 
   -\partial^\mu A^\nu_R \partial^\rho \omega^\sigma  \;\big] \,.
\eea
Here $J^\mu$ is the baryon number current,
given by:
\bea
& & \!\!\!\!\!\!\!
J_\mu  =  \frac{\delta }{\delta \omega^\mu} \Gamma_{WZW}^{\rm full} =
\nonumber \\
& &   \!\!\!\!\!\!\!
-\frac{1}{8\pi^2}\epsilon_{\mu\nu\rho\sigma} 
\big[A^\nu_L \partial^\rho A^\sigma_L - 
A^\nu_R \partial^\rho A^\sigma_R  
 +A^\nu_L \partial^\rho A^\sigma_R  \nonumber \\
& &  \qquad \qquad -A^\nu_R \partial^\rho A^\sigma_L   
 +   A^\nu_L \partial^\rho \omega^\sigma 
   -A^\nu_R \partial^\rho \omega^\sigma \nonumber \\
& &   \qquad \qquad 
- \partial^\nu(\phi/f) (\partial^\rho A^\sigma_L 
+ \partial^\rho A_R^\sigma +2\partial^\rho \omega^\sigma )\nonumber \\
& &  \qquad  \qquad \qquad\qquad
+ \partial^\nu (A^\rho_R \omega^\sigma - A^\rho_L \omega^\sigma)
 \big] \,. 
\eea
where the last two lines contribute zero to the anomaly.

We thus see that the anomalous divergences of currents associated with 
global symmetries are now defined in the 
presence of arbitrary background fields
through the variation of a consistent, gauge-invariant action.  
Note that the baryon number anomaly 
is modified in the presence of the background $\omega$ field
by the $dA d\omega$ terms. Normally, we think of the
global charges and their anomalies
as defined in the limit $\omega\rightarrow 0$,
but we are free to consider the background field corrections
once the gauged currents are defined to be conserved.
In summary: 
{\em The WZW term requires the prescribed counterterm
to recover the correct form of the baryon current anomaly. 
The pCS terms are a consequence of this structure 
and generate new physical interactions.}

From Eq.(\ref{gfull2}) we can anticipate
an interesting new physical application
of anomaly physics in the real world as described in Ref.\cite{HHH2}.
We let $A_L = Z+A$ and $A_R = A$, where $Z$ is the analog of 
the $Z$-boson and $A$ the photon.
Then we obtain from Eq.(\ref{gfull2}) 
the following pCS interaction term:
\begin{multline}
\label{gpcs}
\Gamma_{WZW}^{ \rm full}  =  \Gamma_{WZW} (\phi, A_L, A_R )  
\\
-\frac{1}{8\pi^2}\int  \; \big[\omega (2dA  +   dZ) 
 + \omega d\omega\big]  \left(Z - \frac{d\phi}{f} \right)   \,.
\end{multline}
Gauge invariance of the photon is
manifest, as it must be, since $2dA \equiv F$  
is the electromagnetic field strength.
The $Z$ boson is associated with spontaneous symmetry
breaking.  We see that the gauge shift in $Z$,
$\delta Z = d\epsilon$, is compensated
by $\delta\phi = f \epsilon $ which confirms
gauge invariance 
(again, if a Higgs mechanism is not present to give $Z$ its mass, 
then it will eat the $\phi$ field as in technicolor theories). 

The  interaction in Eq.(\ref{gpcs}) 
is a term in the low-energy effective theory
describing physics at energy scales below the quark mass. 
It contains the massive
gauge field $Z$, and if $M_Z \gg m_q$,  we integrate 
out the $Z$ to derive a set of couplings involving only light fields
as in Ref.~\cite{HHH2}.
The $\omega Z dA$ term leads to a novel
neutrino-photon interaction in nucleons or at
finite baryon density, which may be relevant to
various experiments and astrophysical processes~\cite{HHH2}.

\section{\label{sec:general}The counterterm for general gauging}

The schematic model illustrates a problem that can be posed more
generally as follows.  Consider a ``quark sector'' with a global
(chiral) flavor symmetry $G$ and a subgroup $G'\subset G$ which is
gauged.  In general, $G'$ contains anomalies coming from the quark
sector, so we further assume a lepton sector coupled to the gauge
fields of $G'$, which cancels the quark sector anomalies.  The quarks
are confined, or decoupled, and the flavor symmetry $G$ is broken
spontaneously to a subgroup $H$, giving rise to NGB's that are
elements of the coset space $G/H$.  Some of the NGB's may be eaten by
gauge fields, or the gauge fields may acquire mass from a Higgs
sector.

As a concrete realization of this we can consider the $(u,d)$ quarks
with flavor symmetry $G=SU(2)_L\times SU(2)_R\times U(1)_L\times
U(1)_R$, and spontaneous breaking to $H=SU(2)_{V} \times U(1)_{V}$.
We gauge the $SU(2)_L\times U(1)_Y$ standard model subgroup, and the
$W$ and $Z$ then acquire mass from the usual Higgs boson. The leptons
$(\nu,e)$ will cancel gauge anomalies of the quark sector.

The low energy physics of the quark sector is represented by an
effective lagrangian describing the NGB's and gauge fields $A$.  It
will also contain the spin-1 vector and axial-vector fields, denoted
by $B$ which will be assumed to
transform covariantly under $G'$.  The NGB's are contained in a chiral
matrix field $U$.

Under a general infinitesimal transformation, $\epsilon$, of $G$ we have:
\bea
\label{rep}
\delta U & = & i\epsilon_L(\epsilon) U- iU\epsilon_R(\epsilon) \,,
\nonumber \\
\delta A & = & d\epsilon + i[\epsilon, A] \,,
\nonumber \\
\delta B & = & i[\epsilon, B] \,. 
\eea
Eq.(\ref{rep}) allows for the possibility
of a nonlinear realization, \eg, $\delta U = i\epsilon U -
iU {\epsilon'}(\epsilon, U)$, with 
$\epsilon' \in H$~\cite{Coleman:1969sm}. 
If we specialize to 
$G = U(N_f)_L \times U(N_f)_R$, with   
the associated gauge bosons $A_L$ and $A_R$, 
and background fields $B_L$ and $B_R$
we have:
\beq
\delta U = i\epsilon_L U -
iU{\epsilon}_R \,,
\eeq
and:
\be
\label{transform}
\begin{array}{l}
\delta A_L = d\epsilon_L + i[\epsilon_L, A_L] \,, \\
\delta B_L = i[\epsilon_L, B_L] \,,
\end{array}
\quad 
\begin{array}{l}
\delta A_R = d\epsilon_R + i[\epsilon_R, A_R] \,, \\
\delta B_R = i[\epsilon_R, B_R] \,. 
\end{array} 
\ee

The full effective action contains the kinetic
terms of the NGB's and gauge fields, and any mass
terms associated with explicit breaking (which may
involve the Higgs sector). The effective
action also includes the WZW term, $\Gamma_{WZW}(U, A+B)$,
which represents the anomaly structure of the quark sector.
We also have the contribution, $\Gamma_\ell$, 
to the effective action from
the lepton sector. 

The key point is that the covariant classical background $B$ fields are present 
in the quark sector, but not in the lepton sector.  Mixed terms containing
$A$ and $B$ will thus arise in the gauge anomalies 
of the quark sector, that are 
not cancelled by the lepton sector. 

\subsection{The counterterm}

In deriving the counterterm, we will not need the explicit 
form of the WZW term, but only the consistent anomaly that it generates. 
Consider first the case $B=0$. Then under a
general  gauge transformation in $G'$ we have:
\bea
\label{anom10}
\delta \Gamma_{WZW} & = & -2{\cal C} 
\int \Tr\left[\epsilon\left(dA dA - \frac{i}{2} dA^3\right)\right] \,. 
\eea 
The quantity ${\cal C}$ 
is fixed by properties of the underlying fermion theory. 
For example, for quarks transforming in the fundamental representation of 
$SU(N_c)$, 
\beq\label{Cdef}
{\cal C}= -\frac{N_c}{48\pi^2} \,. 
\eeq 
Eq.(\ref{anom10}) is the
``consistent'' form of the anomaly, before any counterterms are added, 
and it is cancelled by the contribution from the lepton sector:
\bea
\delta \Gamma_\ell & = & 2{\cal C} 
\int\Tr\left[\epsilon\left( dA dA - \frac{i}{2} dA^3\right)\right] \,. 
\eea 
It is convenient to write, modulo a total divergence,
\bea
\delta \Gamma_{WZW} & = & 2{\cal C}
\int \Tr\left[d\epsilon\left(A dA - \frac{i}{2}A^3\right)\right] \,. 
\eea 

Now we introduce the $B$ fields by making the replacement $A\to A+B$ in the
quark sector only. 
This changes the variation of the WZW term so that
under the general gauge transformation
of Eq.(\ref{rep}) we have:
\begin{multline}
\label{anom11f}
\delta (\Gamma_{WZW} + \Gamma_\ell) 
 =  2{\cal C} \int \Tr\bigg\{ 
  d\epsilon \bigg[ 
B dA + dAB + BdB
\\
 - \frac{i}{2}(B A^2 +AB A + A^2B) 
\\
  - \frac{i}{2}(B^2 A +B A B + AB^2)  
-\frac{i}{2}B^3  \bigg] \bigg\} \,. 
\end{multline}
Our problem is to find a counterterm that
cancels this variation. 

The explicit construction of the counterterm
is straightforward, and we obtain the 
result:
\bea
\label{gamc}
\Gamma_c & = & 
-2\C\int \Tr\bigg[ (A dA + dA A)B + \half A (B dB+ dB B) 
 \nonumber \\
& & -\frac{3i}{2}A^3B 
-\frac{3i}{4}AB AB -\frac{i}{2}AB^3 \bigg] \,. 
\eea
The fact that $\delta \Gamma_c = - \delta (\Gamma_{WZW}+ \Gamma_\ell)$
can be verified explicitly.  Therefore, the full action,
\beq
\Gamma = \Gamma_{WZW} + \Gamma_c + \Gamma_{\ell}\,,
\eeq
is now gauge anomaly free in the presence of the NGB's,
gauge fields and spin-1 mesons. 

Note that if we specialize to $G = U(N_f)_L \times U(N_f)_R$, 
with the transformation law (\ref{transform}), 
the counterterm takes the form:
\begin{multline}\label{eq:new}
\Gamma_c  = -2\C \int {\rm Tr}\bigg[
 (A_L dA_L + dA_L A_L) B_L 
\\
+ \frac12 A_L(B_L dB_L + dB_L B_L) 
- {3i\over 2} A_L^3 B_L 
\\
- {3i\over 4} A_L B_L A_L B_L - {i\over 2} A_L B_L^3 \bigg] 
- (L\leftrightarrow R) 
\,. 
\end{multline}

\subsection{Relation to Bardeen counterterm} 

Suppose that we gauge only vector symmetries, for 
example $U(1)_{EM}$ in the standard model. 
We then have: 
\begin{align}
A_L &\to  A + B_L \equiv {\cal A}_L \,, \nl
A_R &\to  A + B_R \equiv {\cal A}_R \,,  
\end{align}
where $B_{L,R}$ are again background fields, transforming covariantly 
under the gauged symmetry. 
The Bardeen counterterm takes the form
$-\Gamma_{WZW}(U=1,{\cal A}_L,{\cal A}_R)$ from Eq.(\ref{WZWfull})
(\eg, see Ref.~\cite{KRS}):
\begin{multline}
\label{bardeenct}
\Gamma_{\rm Bardeen} 
= -\C \int{\rm Tr} \bigg[
 (d\A_R \A_R + \A_R d\A_R ) \A_L 
\\
- (d\A_L \A_L + \A_L d\A_L) \A_R 
\\
-i\left( \A_R^3 \A_L - \A_L^3 \A_R +\frac12 \A_R \A_L \A_R \A_L \right) 
\bigg] \,.
\end{multline}
It can be easily verified that after including the counterterm 
(\ref{bardeenct}), 
the full result $\Gamma_{WZW}+\Gamma_{\rm Bardeen}$ is gauge-invariant
in the vector subgroup.   How does $\Gamma_{\rm Bardeen}$ compare
to our result in Eq.(\ref{eq:new})? 

The Bardeen counterterm 
mixes the $B_L$ and $B_R$ fields.  At first sight, this 
seems to contradict  Eq.(\ref{eq:new}). 
However, upon closer inspection we see 
that all such mixed terms arrange themselves into operators that 
are gauge-invariant in $A$. 
For example, the terms mixing $B_L$ and $B_R$ with one $A$ field and two $B$ 
fields are:
\begin{align}\label{b1}
& \Tr \big[ (dB_R A + dA B_R + B_R dA + A dB_R ) B_L \big]\nl 
- & \Tr \big[ (dB_L A + dA B_L + B_L dA + A dB_L ) B_R \big] \nl
=& \; 3 \Tr \big[ dA (B_R B_L - B_L B_R) \big] \,,
\end{align} 
where a total divergence has been dropped. 
Terms with two $A$'s and two $B$'s are 
\begin{align}\label{b2}
-3i \Tr \big[ A^2 (B_R B_L - B_L B_R) \big]\,.
\end{align} 
Eqs.(\ref{b1}) and (\ref{b2}) combine into the gauge-invariant 
expression,
\be
3\Tr \big[(dA - iA^2) (B_R B_L - B_L B_R ) \big]\,. 
\ee

Continuing in a similar manner, 
the terms mixing $L$ and $R$ are all seen to 
form gauge-invariant 
operators in the fundamental gauge fields $A$. 

Splitting the Bardeen counterterm into a gauge invariant
and anomalous piece,
\beq
\Gamma_{\rm Bardeen} = \Gamma_{\rm Bardeen}^{\rm G.I.} 
+ \Gamma_{\rm Bardeen}^{\rm anom.} \,, 
\eeq
we find:
\begin{multline}
\Gamma_{\rm Bardeen}^{\rm G.I.} = 
{-\C}\int{\rm Tr}\bigg\{ 
3(dA-iA^2) (B_R B_L - B_L B_R) 
\\
+ (DB_R B_R + B_R DB_R) B_L 
- (DB_L B_L + B_L DB_L) B_R 
\bigg\} \,. 
\end{multline}
Here the covariant derivatives acting on $B_{L,R}$, taking account 
of the anticommuting forms, are 
\begin{align}
DB_L &= dB_L - iAB_L - iB_LA \,, \nl
DB_R &= dB_R - iAB_R - iB_RA \,. 
\end{align}
The remaining, anomalous part 
can be simplified to:
\begin{align}
\label{bc_final}
\Gamma_{\rm Bardeen}^{\rm anom.} &= \Gamma_c =
-\C\int{\rm Tr}\bigg[  \nl
& \!\!\!\!\! \!\!\!\!\! \!\!\!\!\! \!\!\!\!\! \!\!\!\!\! 
2(dA A + A dA) B_L + A(dB_L B_L + B_L dB_L) \nl 
& \!\!\!\!\! \!\!\!\!\! \!\!\!\!\! \!\!\!\!\! \!\!\!\!\! \!\!\!\!\!
-i\left( 3A^3 B_L + \frac32 AB_L AB_L + AB_L^3 \right)
\bigg] - (L \leftrightarrow R)  \,. 
\end{align}
When $A_L = A_R = A$, \ie, when only vector symmetries are gauged,
our general expression Eq.(\ref{eq:new}) reduces to precisely this form.
Our new counterterm is the 
generalization of the Bardeen counterterm 
when the gauge subgroup $G'$ is not
contained in the unbroken subgroup $H$ of the chiral theory $G/H$. 

The Bardeen counterterm has been well-studied in the past. 
For example, in Ref.~\cite{Chu:1996fr}, the Bardeen counterterm is 
employed for the purpose of constructing the gauged 
WZW term by an integration formula that requires vanishing anomaly 
in the unbroken ($H$) subgroup of a general $G/H$.   
Inclusion of the Bardeen counterterm can be phrased as the 
boundary condition $\Gamma(U=1) = 0$ when integrating the 
anomaly~\cite{Wess}.  
Our results show, however, that this boundary condition is incompatible with 
gauge invariance in the general case involving non-vectorlike gauging. 

The Bardeen counterterm has also appeared in phenomenological 
analyses~\cite{KRS}. 
However, such analyses neglect the important effects of neutral and 
charged weak currents, and have added the counterterm in an {\it ad hoc} 
manner: 
only photon gauge invariance is preserved, 
and global chiral symmetries are broken even in the absence of gauge 
fields. 
The Bardeen counterterm maintains 
gauge invariance in the presence 
of background fields {\em only when vector symmetries are gauged. }
It is not the appropriate construct 
when the full
standard model $SU(2)_L\times U(1)_Y$ gauging is relevant.

\section{Global current anomalies and gauge invariant 
operators \label{sec:cov}}

The new counterterm is necessary for a 
proper derivation of global current 
anomalies, such as 
the baryon current anomaly 
in the standard model.  
Our counterterm ensures that the  action is anomaly
free under the gauged $G'$ symmetry,   
in the presence of arbitrary background 
fields.  This action still has a number of global symmetries 
that are not broken explicitly by gauging, namely 
the special transformations for which $[\epsilon, A]=0$.  
The associated symmetry currents are generated by varying the
background fields, and are conserved modulo anomalies.  
Since our theory is locally gauge invariant under
$G'$ transformations, the global anomalies generated from
the full action will automatically be gauge covariant expressions
in the $A$.    In this sense, they are "covariant" anomalies.
The formalism also implies that these currents and
anomalies necessarily contain the background spin-1 meson
fields, $B$.
Note that since 
the global anomalies are derived from a well-defined action,
they necessarily satisfy the appropriate extension of Wess-Zumino
consistency conditions that describes variations with respect to
both $A$ and $B$ fields.

\subsection{The general case}

Let us then consider the variation,
\bea
\label{rep2}
\delta U & = & i\epsilon_L(\epsilon) U- iU\epsilon_R(\epsilon) \,,
\nonumber \\
\delta A & = &  i[\epsilon, A] = 0 \,,
\nonumber \\
\delta B & = & d\epsilon + i[\epsilon, B] \,. 
\eea
Since $B$ enters only the ``quark'' sector, 
and we impose $[\epsilon, A]=0$,  the lepton 
effective action, $\Gamma_\ell$, is
invariant. We therefore need
consider only the variation of 
$\Gamma_{WZW}(U,A+B)+ \Gamma_c(A,B)$. 

We thus obtain the general expression for the global anomaly:

\begin{multline}
\delta (\Gamma_{WZW} + {\Gamma}_c ) =  
 -2\C
\int\Tr\bigg\{ \epsilon\bigg[ 
\\
3(dA-iA^2)^2  + 3(dA-iA^2) DB 
\\
+ (DB)^2 -\frac{i}{2}D(B^3)
+ iB(dA-iA^2)B  
-{i}(dA-iA^2)B^2\bigg]\bigg\} \,,
\end{multline}
where $DB = dB-iAB-iBA$.
Note the appearance of the 
covariant field strength, $(dA-iA^2)$.
We emphasize that the form of this
result depends on the condition $[\epsilon,A]=0$. 
In the explicit chiral representation for $U(N_f)_L\times U(N_f)_R$
the anomaly takes the form:
\bea
\label{resglo}
&& 
\!\!\!  \!\!\!  \!\!\!  
\delta (\Gamma_{WZW} +{\Gamma}_c ) = \nl
&&
\!\!\!  \!\!\!  \!\!\!  
 -2\C
\int\Tr\bigg\{ 
\epsilon_L\bigg[ 3(dA_L-iA_L^2)^2  
+ 3(dA_L-iA_L^2)(DB_L) 
\nl
&& \quad 
+ DB_LDB_L -\frac{i}{2}D(B_L^3)
+ iB_L(dA_L-iA_L^2)B_L  
\nl
&& \quad 
-{i}(dA_L-iA_L^2)B_L^2\bigg]
\bigg\} \;\; - \;\;
 (L\leftrightarrow R) \,,
\eea
with $DB_L = dB_L -iA_LB_L -i B_LA_L$.

\subsection{Application to the Standard Model}

Let us illustrate the computation of covariant anomalies by
considering the baryon current of the first generation quarks in 
the standard model.  
We will first give a description at the quark level, emphasizing
that the counterterm is required for a correct derivation of
the anomaly.   We then give an equivalent description at 
the chiral lagrangian level.  This leads to a generalization of the 
Goldstone-Wilczek current in the presence of gauge fields. 

Let $Q=(u,d)$ and consider the action:
\begin{multline}
\label{baction}
S_Q = \int d^4x\, \bar{Q}_L(i\slash{\partial} + \slash{A}_L + \slash{B}_L )Q_L 
\\
+ \bar{Q}_R(i\slash{\partial} + \slash{A}_R + \slash{B}_R )Q_R \,,
\end{multline}
where $A_{L\mu} = g_2W_\mu^a \tau^a/2 + g_1 W^0_\mu Y_L/2$, 
$A_{R\mu}=g_1 W^0_\mu Y_R/2$, 
$B_L = B_R = \omega_\mu \,diag(1/3,1/3)$. 
Under variation in $\omega_\mu$ we obtain
the baryon current:
\beq\label{bcur}
\frac{\delta S_Q}{\delta\omega_\mu} = J^\mu 
= \frac{1}{3}\bar{Q}\gamma^\mu Q \,. 
\eeq

By considering the local variation 
$\delta\omega = d \epsilon$, 
we obtain from the
Weyl quark loops the consistent anomaly of the quark effective 
action: 
\bea
\label{bdiv}
&& \delta S_Q \big|_{\omega = 0} = \nonumber \\ 
&& -2\C\int \, \epsilon \, \Tr\left\{ 
\frac{1}{3}\left[(dA_L)^2 -
\frac{i}{2}d(A_L^3)\right]\right\} -(L\leftrightarrow R) \,.
\nonumber \\ 
\eea
We must also include 
the counterterm Eq.(\ref{eq:new}), which 
takes the form, to leading order in $\omega$:
\bea
\label{ctct}
&& \Gamma_c   =  \nl
&&
2\C \int {\rm Tr} \bigg\{ 
\frac{1}{3}\omega\bigg[ 
A_L dA_L + dA_L A_L - {3i\over 2} A_L^3\bigg] \bigg\} -(L\leftrightarrow R) 
\,. 
\nl
\eea
Under the local variation $\delta\omega = d\epsilon$ we find:
\bea
&& \delta\Gamma_c  = \nonumber \\
&& -2\C \int {\rm Tr} \bigg\{ 
\frac{1}{3}\epsilon\left[ 
2dA_L dA_L - {3i\over 2} d(A_L^3)\right] \bigg\} -(L\leftrightarrow R),  
\nonumber \\ 
\eea
and upon combining the quark loop
contribution with the counterterm we find:
\begin{align}
& \left. \delta S_Q  + \delta\Gamma_c \right|_{\omega = 0} =
\nonumber \\
& -2\C\int\Tr\left[\epsilon(dA_L - iA_L^2)^2
\right]-(L\leftrightarrow R)  
\nonumber \\  
&
= \int d^4x\, \epsilon \,
\frac{1}{64\pi^2}\epsilon_{\mu\nu\rho\sigma} \left(
g_2^2 F^a_{\mu\nu} F^a_{\rho\sigma}
+ \half\Tr{Y^2} g_1^2 F^Y_{\mu\nu} F^Y_{\rho\sigma}\right) \,,
\end{align}
where $F^a_{\mu\nu} = \partial_\mu W^a_\nu-\partial_\nu W^a_\mu 
+ g_2 \epsilon^{abc}W_\mu^b W_\nu^c$ is the covariant $SU(2)_L$ field
strength, and $F^Y_{\mu\nu} = \partial_\mu W^0_\nu-\partial_\nu W^0_\mu $
is the weak hypercharge field strength. 
The factor $\Tr(Y^2) = 2\times(1/3)^2
- (4/3)^2 - (-2/3)^2 = -2$ is traced over the $(u,d)$ quarks. 
Note that we could also have read this result directly from Eq.(\ref{resglo}).
 
Hence, with 
$\tilde{F}_{\mu\nu}=(1/2)\epsilon_{\mu\nu\rho\sigma} F^{\rho\sigma}$:
\beq
\label{Banom}
\partial_\mu J^\mu = -\frac{1}{32\pi^2}(g_2^2 F^a_{\mu\nu} \tilde{F}^{a\mu\nu}
- g_1^2F^Y_{\mu\nu} \tilde{F}^{Y\mu\nu}) \,.
\eeq
 While this result can be obtained by
naively rescaling Feynman diagrams, using Adler's axial vector anomaly
from QED as a starting point,
the result would then be only fortuitously correct; the modified counterterm
structure is required to generate the
formally correct baryon current anomaly in the standard model. 
For $B\pm L$ we need a similar counterterm construction
in the lepton sector with an auxiliary background
field, \etc.  Of course, the $B-L$ anomaly
cancels between leptons and quarks, insofar as we
take the limit of zero background fields after 
calculating the current divergence.
The $B+L$ anomaly is $2$ times 
the above result, Eq.(\ref{Banom}).

Note that if we had inadvertently used the Bardeen counterterm
of Eq.(\ref{bardeenct}) in defining the baryon number current and
divergence, we would have:
\begin{multline}
\Gamma_{\rm Bardeen}  = 
-2\C \int {\rm Tr} \bigg[ 
\frac{1}{3}\omega\bigg( 
A_L dA_L -  {i\over 2} A_L^3 + 3 dA_R A_L \\
- A_R dA_R + {i\over 2} A_R^3 - 3 dA_L A_R 
\bigg) \bigg] \nonumber \,,
\end{multline}
and a short calculation shows that 
the resulting baryon number current would have {\it zero}
divergence in place of $1/32\pi^2$ in Eq.(\ref{Banom}).     

\subsection{Generalization of the Goldstone-Wilczek Current}

We can give an equivalent description of the anomalies at 
the level of the chiral theory of mesons, instead of at the quark level.
The WZW term, with the new counterterm, can be
expanded in the external fields $B$:
\beq
\label{meth}
\Gamma(A,B,U) = \Gamma(A,U) + \int d^4x\, 
{\rm Tr}(B_\mu J^\mu) + {\cal{O}}(B^2) \,. 
\eeq
The leading term in the expansion, $\Gamma(A,U)$, 
is the original WZW term.  It generates the consistent gauge anomaly,
which is cancelled by the leptons. Therefore the subsequent
terms, such as $\Tr(B_\mu J^\mu)$, must be gauge invariant. 
In particular, $J_\mu$
is the covariant global current associated with $B$.

Let us again focus on the baryon number current
in the $U(2)_L\times U(2)_R$ 
chiral theory, so that 
$B_L=B_R = \omega \; diag(1/3,1/3)$.
First note that, in the limit
$A_L=A_R = 0$ we see, from the WZW term reproduced below 
in Eq.(\ref{WZWfull}), that: ($\alpha = dU U^\dagger$)
\bea
J_\mu & = & \frac{\delta }{\delta \omega_\mu}\Gamma_{WZW} =
\frac{2}{3}\C\epsilon_{\mu\nu\rho\sigma}\Tr(\alpha^\nu\alpha^\rho\alpha^\sigma ) 
\nonumber \\
& = & \frac{N_c}{72\pi^2}\epsilon^{\mu\nu\rho\sigma}\Tr
(U\partial^\nu U^\dagger  U\partial^\rho U^\dagger U\partial^\sigma U^\dagger )
\,. 
\eea
This is the Goldstone-Wilczek topological current that
describes baryon number in the chiral lagrangian~\cite{GW}. 
The current
arises automatically upon introducing the background $\omega$
field into the WZW term. With $N_c=3$ it yields
a baryon number of $1$ for a Skyrmion hedgehog configuration.

Armed with our new counterterm and using Eqs.(\ref{meth},\ref{WZWfull}) we
can compute the form of the 
baryon current in the presence of the gauge fields. 
For simplicity we keep just the $SU(2)_L$ part.
The result is a gauge invariant current,
as it must be, and for $N_c=3$:  
\beq
J_\mu = -\frac{1}{24\pi^2}\epsilon^{\mu\nu\rho\sigma}\Tr
\left(
\tilde{\alpha}_\nu \tilde{\alpha}_\rho \tilde{\alpha}_\sigma 
+ \frac{3i}{2} F_{L\nu\rho}\tilde{\alpha}_\sigma  
\right) \,.
\eeq
Here:
\begin{align}
\tilde{\alpha}_\mu & = (D_\mu U) U^\dagger 
\,, \nl
D_\mu
&= \partial_\mu -iA_{L\mu}  
\,, \nl
F_{L \mu\nu} &= {i}[D_\mu, D_\nu] \,. 
\end{align}
This current reduces to the Goldstone-Wilczek result when $A_L\rightarrow 0$.

A similar current using electromagnetic gauging
was constructed by Callan and Witten, where the new term 
is seen to play
a crucial role in monopole catalysis of baryon number
viewed at the Skyrmion level \cite{Witten2}.
The nonabelian form was constructed 
using dimensional deconstruction, matching
to a Yang-Mills topological current in $D=5$  
in Ref.~\cite{hillc}.
The present derivation by variation of an action is more 
general, and it is now straightforward to construct any of the
chiral topological currents by variation of the WZW term plus counterterm.
This nonabelian $SU(2)_L$ current can arise only when we use the WZW term 
with our improved counterterm $\Gamma_c$.

Note that we can compute explicitly, for $N_c=3$,
\beq
\label{anom33}
\partial_\mu J^\mu = -\frac{1}{16\pi^2}\Tr(F_{L\mu\nu}\tilde{F}_L^{\mu\nu}) \,,
\eeq
where  a useful identity (the Bianchi identity for a
deconstructed $D=5$ theory \cite{cth}) is:
\bea
\label{alpha1}
[ D_\mu, \tilde{\alpha}_\nu ] - [D_\nu, \tilde{\alpha}_\mu]
& = &
[\tilde{\alpha}_\mu, \tilde{\alpha}_\nu] -iF_{L\mu\nu}  \,. 
\eea
Eq.(\ref{anom33}) reproduces the result for $SU(2)_L$ 
obtained 
in Eq.(\ref{Banom}).  
Note that for a self-dual instanton in which $F_{\mu\nu} = \tilde{F}_{\mu\nu}$
and we have the Euclidean action $(1/2 g_2^2) \int \Tr{FF}= 8\pi^2 /g_2^2 $,
the baryon charge is changed by one unit, and $B+L$ changes by $2$ units,
confirming the usual intuition. 

It is interesting to contemplate the full background
field ($B$) structure of
Eq.(\ref{resglo}).  Note that the $B$-field containing
terms are a total divergence, and can be absorbed
by a redefinition of the baryon current. However,
they probably do have a physical role to play at high baryon density.
Note that there is no $F^Y d\omega$ term 
in the baryon current anomaly, 
owing to ${\rm Tr} (Y_L-Y_R) = 0$.  However,
if we keep the $\rho$ meson then there are surviving mixed isospin
and weak-isospin
terms $\Tr(F_W d\rho)$ ($F_W$ is weak isospin).  
Whether there is more to this story, \eg, an 
enhancement of baryon number violation at large finite baryon 
density through
this form of the mixed anomaly, or a description
of certain superfluid phases of baryons, remains
to be investigated.  

\subsection{Gauge invariant operators}

Throughout this discussion, we have implicitly assumed that
the counterterm (\ref{gamc}) is unique.   
In fact, we can construct additional 
gauge invariant operators beginning at 
quadratic order in the $B$ fields.  
While in the limit $B\rightarrow 0$
the gauge field part of the global anomaly is
uniquely determined, these terms can potentially 
lead to an ambiguity in the structure of the global anomaly 
at finite $B$.     
For simple group models, we can add a general counterterm 
of the form,
\be
\Gamma_{\rm c,\,G.I.} = \int 
c_{1} 
{\rm Tr}\big[ 
B^2 (dA - iA^2) 
\big] \,, 
\ee
with $c_1$ a free parameter.  In fact, 
it is easy to show that
the general anomaly (\ref{resglo}) is not affected by this term.

For product group models, the situation is slightly more complicated. 
A short calculation, after dropping total derivatives, yields the 
general expression for term containing at least one $A$ field:
\begin{align}\label{g.i.}
\Gamma_{\rm c,\,G.I.} &= 
{N_c\over 24\pi^2}
\int\bigg\{ 
c_{1L} 
{\rm Tr}\big[ 
B_L^2 (dA_L - iA_L^2) 
\big]\nl
&\quad 
+ c_{2L} 
{\rm Tr}\big( B_L ) 
{\rm Tr}\big[ B_R (dA_R - iA_R^2 ) \big] \nl
&\quad 
+ c_{3L}
{\rm Tr}\big( B_L )
{\rm Tr}\big[ B_R DB_R \big] 
 \bigg\} 
+ 
(L \leftrightarrow R) 
\,. 
\end{align}
To recover a parity-symmetric theory when only vector symmetries are 
gauged, we should have $c_L=-c_R$.   
The effects of these operators on the anomalies at finite $B_{L,R}$ 
can be worked out in the general 
case.  For example, 
when the field $\omega$ coupling to baryon number is the only 
background field present, then $\Gamma_{c, \rm G.I.}=0$, since 
$\epsilon^{\mu\nu\rho\sigma}\omega_\mu \omega_\nu \dots =0$.  

\section{Pseudo-Chern-Simons terms for the Standard Model\label{sec:sm}}

We can apply the results from the previous section to
compute the explicit form of the pCS terms for the standard model.  
For simplicity, we focus on a single standard model generation,
\ie, the $(u,d)$ quarks and the $(\nu_e, e)$  leptons. 
The low energy physics of the quark sector is represented by 
a $U(2)_L\times U(2)_R$
chiral lagrangian describing interactions of the three pions and the
$\eta$, 
gauge fields, and vector mesons.  
The lepton sector is also present in the effective theory.
We introduce the full $SU(2)_L\times U(1)_Y$ gauging and
gauge anomalies cancel between the quark and lepton sector.
We also include the spin-1 vector mesons, treated as classical background
fields, corresponding to the 
$\rho^{0,\pm}$, $\omega^0$ and $a_1^{0,\pm}$, $f_1^0$ vector
and axial-vector mesons. 
At $\lta $ GeV  energies, where the chiral lagrangian 
description is appropriate, the $W$ and $Z$ bosons may be 
subsequently integrated
out of the theory, and their effects represented by the corresponding
charged and neutral weak currents.

The complete effective lagrangian thus contains the kinetic
terms of the NGB's, leptons, gauge and vector meson fields, 
and any mass terms associated with symmetry breaking (which may
involve the Higgs sector for the $SU(2)_L\times U(1)_Y$
breaking).  
The effective action also includes the WZW term and 
the counterterm:
\beq
\Gamma_{WZW}^{\rm full} = \Gamma_{WZW}(U, A+B)+ \Gamma_c(A,B),
\eeq
which represent the full anomaly structure of the quark sector.

For the fundamental gauge fields we write
\begin{align}
A_L &= g_2 W^a{\tau^a/2} + g_1 W^0 \left( 
\begin{array}{cc} \frac16 \\ & \frac16 \end{array}
\right) \,, \nl  
A_R &= g_1 W^0 \left( 
\begin{array}{cc} \frac23 \\ & -\frac13 \end{array}
\right) \,. 
\end{align} 
(we use $W^0$ to denote the $U(1)_Y$ gauge field, so
as not to confuse with our previous usage of $B$ as a generic classical
background field).
In terms of the charge and 
mass eigenstates after electroweak symmetry breaking we have:
\begin{align}
W^0 &= -s_W Z + c_W A \nl
W^1 &= {1\over \sqrt{2}} W^+ + {1\over\sqrt{2}} W^- \nl
W^2 &= {i\over \sqrt{2}} W^+ - {i\over\sqrt{2}} W^- \nl
W^3 &= c_W Z + s_W A \,, 
\end{align}
where $c_W = g_2/\sqrt{g_1^2+g_2^2}$, $s_W=g_1/\sqrt{g_1^2+g_2^2}$.   

Let us now put the standard model 
in the classical background of vector and axial-vector 
mesons:
\begin{align}
B_V &\equiv B_L + B_R = 
g
\left(\begin{array}{cc} 
\rho^0 & \sqrt{2} \rho^+ \\
\sqrt{2} \rho^- & -\rho^0 
\end{array}
\right) 
+ g'
\left(\begin{array}{cc} 
\omega &  \\
 & \omega 
\end{array}
\right) 
\,, \nl
B_A &\equiv B_L - B_R = 
g
\left(\begin{array}{cc} 
a^0 & \sqrt{2} a^+ \\
\sqrt{2} a^- & -a^0
\end{array}
\right) 
+ g'
\left(\begin{array}{cc} 
f & \\
 & f
\end{array}
\right) \,. 
\end{align}
where $\rho$ and $a$ are isotriplets and $\omega$ and $f$ are iso-singlets. 
Note the slightly unconventional definitions $V,A = (A_L\pm A_R)$ instead of 
$V,A = (A_L \pm A_R)/2$; the resulting normalization of 
$g$, $g'$ can be more readily compared to the literature.   
Note also that this normalization implies that
\be
g_\omega = \frac{3}{2} g' \,,
\ee
where $g_\omega$ multiplies the field that is coupled to 
baryon number~\cite{HHH1}.
In the following, we suppress coupling constants for the
vector fields.  
We can recover the complete result with couplings by taking
$(A,W,Z)\to (g_2 A,g_2 W,g_2Z)$, 
$(\rho,a)\to (g\rho,ga)$, $(\omega,f)\to (g'\omega,g'f)$.   
 
\subsection{Fundamental gauge fields and anomaly cancellation}

The WZW term for $U(N_f)_L \times U(N_f)_R \to U(N_f)_{V}$ 
is given in a convenient 
form  by Kaymakcalan, Rajeev
and Schechter \cite{KRS}%
\footnote{We have the opposite overall
sign to \cite{KRS} because, given $D= d-iA$ 
and interpreting $\Gamma_{WZW}$ as an {\em action}
arising from
the quark level action $\int \bar{q}i\slash{D}q = \int A \cdot J + ...$
it must shift by $-\int \epsilon \partial\cdot J $ 
under a gauge transformation. Our choice of the sign of $\C$ ensures this.
}.
In terms of: 
\be
\A_{L,R} = A_{L,R} + B_{L,R} \,, \qquad \C = -\frac{N_c}{48\pi^2},
\ee
we have: 
\begin{widetext}
\begin{align}
\label{WZWfull}
&{\Gamma}_{WZW}(U, \A_L, \A_R)  = \Gamma_0(U) \;  + {\cal C} \int \Tr \bigg\{  
({\A}_L\alpha^3 + {\A}_R\beta^3 ) 
-\frac{i}{2}[({\A}_L\alpha)^2 - ({\A}_R\beta)^2 ] 
\nonumber \\
&\quad
+i(\A_L U {\A}_R U^\dagger \alpha^2 - {\A}_R U^\dagger \A_L U \beta^2)  
+i ( d\A_R dU^\dagger \A_L U - d\A_L dU \A_R U^\dagger  ) 
\nonumber \\
&\quad
+i \big[ (d\A_L\A_L + \A_L d\A_L)\alpha+ (d{\A}_R{\A}_R +{\A}_Rd\A_R)\beta \big]
+({\A}^3_L \alpha+{\A}^3_R \beta)
\nonumber \\
&\quad
- (d\A_L \A_L + \A_L d\A_L )U \A_R U^\dagger +(d\A_R \A_R + \A_R d \A_R) U^\dagger \A_L U
\nonumber \\
&\quad
+(\A_L U  \A_R U^\dagger \A_L \alpha + \A_R U^\dagger \A_L U \A_R \beta  ) 
+i\left[ \A_L^3 U  \A_R U^\dagger - \A_R^3 U^\dagger \A_LU - 
\half (U\A_RU^\dagger \A_L)^2 \right]
\bigg\} \,. 
\end{align}
Here $\alpha = dU U^\dagger$ and $\beta =
U^\dagger dU$.  
The function $\Gamma_0$ is given by  
\begin{equation}\label{Gamma0}
{\Gamma}_{0}(U)  = -{i {\cal C} \over 5} \int_{M^5} \Tr\left( \alpha^5\right)
= \frac{i N_c}{240\pi^2}\int d^5x \,  \epsilon^{ABCDE}\;
\Tr \left( \alpha_A\alpha_B\alpha_C\alpha_D\alpha_E \right) \,, 
\end{equation}
\end{widetext}
where $M^5$ is a five dimensional manifold with spacetime as its boundary. 
The quantization condition ensures that $e^{i\Gamma_0}$ is independent of 
the choice of bounding surface. 
In four dimensions, 
\begin{equation}
{\Gamma}_{0}(U) = 
-{2N_c\over 15\pi^2 f_\pi^5} \int_{M^4} {\rm Tr}\left[ \pi (d\pi)^4 \right] 
+ ...  \,. 
\end{equation}

Also, we have the new counterterm $\Gamma_c$, which is given by
Eq.(\ref{eq:new}).
We now take $\Gamma_{WZW}^{\rm full} = \Gamma_{WZW} + \Gamma_c$ and examine
the terms containing mixed factors of fundamental
gauge fields, $A$'s, and classical background fields, $B$'s.
Notationally, in the following, \eg, a term denoted $\Gamma_{AABB}$ contains 
two factors of $A$ and two factors of $B$, \etc.    

The terms in $\Gamma_{WZW}^{\rm full}$ involving 
just the fundamental gauge fields $A_{L,R}$  do
not involve the counterterm.  Explicitly, we see that 
the terms with three and four fundamental fields read:  
\begin{widetext}
\bea
\label{aaa}
\Gamma_{AAA} &=&
{\cal C} \int 
dZ A Z \left(\frac23 s_W - \frac12 {s_W\over c_W^2}\right) 
+ dA A Z \left( \frac23 {s_W^2\over c_W} \right)  
+ (dW^+ W^- + dW^- W^+ )\left( \frac16 {s_W^2\over c_W} Z 
- \frac16 s_W A \right) \,, 
\eea
\end{widetext}
\bea
\label{aaaa}
\Gamma_{AAAA} &=& {\cal C} \int 
i W^+ W^- Z A \left( \frac14 {s_W\over c_W} \right) \,. 
\eea
The terms $\Gamma_{AAA}$ and $\Gamma_{AAAA}$ combine with the
lepton-sector loop contributions to produce 
gauge invariant operators that do not 
involve the vector meson fields.  
In a formal limit where we assume the leptons are heavy, we can integrate 
out $(\nu,e)$ to 
obtain the lepton-sector WZW term as a function of $W$, $Z$, $\gamma$ 
and the NGB's of the Higgs boson. 
In this case, the pure gauge terms Eqs.(\ref{aaa},\ref{aaaa}) 
cancel exactly against 
corresponding lepton sector loop contributions~\cite{dhokerII,rjh}.

\subsection{Interactions involving vector meson fields}

Since the $B$ fields transform linearly under the gauge transformations, 
the sum of the remaining terms must be separately gauge invariant. 
For the various remaining terms in $\Gamma_{WZW}^{\rm full}$ the result is: 
\begin{widetext}
\bea
\label{smcounter}
\Gamma_{AAB} &=& 
{\cal C} \int 
dZ Z \left[ {s_W^2\over c_W^2} \rho^0 + \left({3 \over 2 c_W^2}-3\right)\omega 
-{1\over 2 c_W^2} f \right] 
+ dA Z \left[ -{s_W\over c_W} \rho^0 - {3 s_W\over c_W} \omega \right] 
+ dZ \left[ W^- \rho^+ + W^+ \rho^- \right]{s_W^2\over c_W} 
\nl
&&\qquad
-s_W dA \left[ W^- \rho^+ + W^+ \rho^- \right] 
+(DW^+ W^- + DW^- W^+)\left[ -\frac32 \omega -\frac12 f \right]  \,, 
\nl
\Gamma_{ABB} &=& 
{\cal C} \int
Z \bigg\{ d\rho^0 \bigg[ -{3\over 2c_W}\omega -{s_W^2\over c_W} a^0 
+ \left( -{3\over 2 c_W} + 3 c_W \right) f \bigg]
+ d\omega \bigg[ -{3\over 2 c_W}\rho^0 + \left( -{3\over 2 c_W} + 3 c_W \right)a^0
-{s_W^2\over c_W} f \bigg] \nl
&& \quad 
+ da^0 \bigg[ {s_W^2\over c_W}\rho^0 + \left({3\over 2 c_W} -3 c_W\right)\omega
-{1\over 2 c_W} f \bigg] 
+ df \bigg[ \left( {3\over 2 c_W} -3 c_W \right)\rho^0 + {s_W^2\over c_W} \omega
- {1\over 2c_W} a^0 \bigg] \bigg\} 
\nl
&& \quad 
+ s_W dA \left( \rho^0 a^0 + 3\rho^0 f + 3\omega a^0 + \omega f 
+ \rho^+ a^- + \rho^- a^+ 
\right)  
-{s_W^2\over c_W} dZ \left( \rho^+ a^- + \rho^- a^+ \right) 
\nl
&& \quad 
+ \frac32 \left[W^+ D\rho^- + W^- D\rho^+) \right] \left(-\omega + f\right) 
+ \frac32 \left[ W^+ (-\rho^- + a^-) + W^- (-\rho^+ + a^+) \right]d\omega \nl
&& \quad 
+ \frac12 \left[ W^+ Da^- + W^- Da^+ \right](-3\omega - f) 
+ \frac12 \left[ W^+ (-3\rho^- -a^-) + W^- (-3\rho^+ - a^+) \right] df \,, 
\nl 
\Gamma_{BBB} &=& {\cal C}\int
2\bigg[ \left( \rho^- f + \omega a^- \right) D\rho^+  
+ \left(\omega a^+ + \rho^+ f\right)D\rho^- 
+ \left( \omega a^0 + \rho^0 f \right) d\rho^0 
+ \left( \rho^+ a^- + \rho^- a^+ + \omega f + \rho^0 a^0 \right) d\omega 
\bigg]
\,,
\nl
\Gamma_{AAAB} &=& {\cal C}\int 
i 
W^+ W^- Z \left[ 3c_W \omega 
+  \left(c_W + {1\over 2 c_W}\right) f \right] \,,
\nl 
\Gamma_{AABB} &=& {\cal C}\int i\bigg\{ 
W^+ W^- \bigg[ \frac32 (\rho^0+a^0) \omega -\frac12(\rho^0-a^0) f \bigg] 
\nl
&& \quad 
+ W^+ Z \bigg[ \left( {3c_W\over 2} - {1\over c_W}\right)\rho^- f - {3c_W\over 2} \rho^- \omega -{c_W\over 2} a^- f 
+{3c_W\over 2} \omega a^- \bigg] \nl
&& \quad
+ W^- Z \bigg[ \left(-{3c_W\over 2} + {1\over c_W}\right)\rho^+ f 
+ {3c_W\over 2} \rho^+ \omega + {c_W\over 2} a^+ f - {3c_W\over 2}
\omega a^+  \bigg] 
\bigg\} \,,
\nl 
\Gamma_{ABBB} &=& {\cal C}\int i\bigg\{ 
W^+\bigg[ \rho^-\rho^0(\omega-2f) -\rho^- \omega a^0 +\rho^0 \omega a^- + \omega a^- a^0 \bigg] 
+W^- \bigg[ \rho^+\rho^0(-\omega + 2f) +\rho^+\omega a^0 -\rho^0\omega a^+ -\omega a^+ a^0 \bigg] \nl
&&\quad
+ Z \bigg[ {1\over c_W} \rho^+ \rho^- \omega + \left(-2c_W +{1\over c_W} \right) 
\left( 2 \rho^+ \rho^- f +  \rho^+ \omega a^- - \rho^- \omega a^+ \right)
+ {1\over c_W} \omega a^+ a^-  \bigg] 
\bigg\} 
\,. 
\eea
\end{widetext}
These results use the abbreviated notation of
differential forms, so that for example 
$\int d^4x\; 
\epsilon_{\mu\nu\rho\sigma}A^\mu B^\nu \partial^\rho C^\sigma = \int
ABdC$.
Here we have defined covariant derivatives of the charged fields as 
\bea
D W^\pm &=& dW^\pm \mp is_W A W^\pm \,, \nl
D \rho^\pm &=& d\rho^\pm \mp is_W A \rho^\pm \,, \nl
D a^\pm &=& da^\pm \mp is_W A a^\pm \,. 
\eea
Note that the photon always appear as a field strength, 
or as a covariant derivative acting on 
charged vector bosons or mesons, as required by gauge invariance. 
Gauge invariance in $W$ and $Z$ is not explicit, since we have not included
terms involving the pion fields.  

Some notable interactions in the above include the term from $\Gamma_{AAB}$:  
\be\label{newint}
- {3s_W\over c_W} g_2^2 g' \C \int \omega Z dA \,.
\ee
This interaction was studied in Ref.\cite{HHH2} and 
mediates neutrino-photon interactions in nuclear matter.   
%
It should be noted that without including 
$\omega$ in the WZW term, 
the $\omega Z dA$ interaction can still be obtained
from the assumption that the physical $\omega$ couples
to the baryon current, $J_\mu$, through a phenomenological interaction 
of the form $g_\omega \omega^\mu J_\mu$.
At low energies we must then use for $J_\mu$  
the modified form of the Goldstone-Wilczek current 
in the presence of $Z$ and $A$ as dictated by gauge invariance
and the new counterterm.  This contains 
the $\omega Z dA$ interaction precisely as in Eq.(\ref{newint}).

Note that in $\Gamma_{ABB}$ we find a term:
\be
3 s_W g_2 g g' {\cal C} \int dA \rho^0 f \,. 
\ee
This term mediates the decay $f_1 \to \rho^0 \gamma$ and is
in reasonable accord with experiment.   
We will study such interactions in more detail elsewhere~\cite{HHH1}.  

In $\Gamma_{BBB}$ we find 
\be
2 g^2 g' 
{\cal C} \int (\omega a^0 d\rho^0 + \rho^0 a^0 d\omega ) \,. 
\ee
Operators of this form have been studied in Ref.~\cite{Domokos}. 
However, this term can be modified by 
the gauge invariant operators of Eq.(\ref{g.i.})   
whose coefficients are not fixed solely by anomaly matching arguments.    

\section{\label{sec:conc}Conclusions}

In this paper we constructed the gauged WZW term for  the
standard model $SU(2)_L \times U(1)_Y$ gauge subgroup of the $U(2)_L \times
U(2)_R$ chiral symmetry of the strong interactions. An essential
ingredient in this construction was the demand that we maintain gauge
invariance in the presence of background fields coupled to the $U(2)_L
\times U(2)_R$ currents.

These background fields play two distinct but equally important
roles. First, they allow us to define the currents of global
symmetries through variation of the background fields.  When applied
to the (anomalous) baryon current this allowed us to derive a
generalization of the Goldstone-Wilczek current in the presence of
both fundamental gauge fields and background vector fields.  Second,
in environments where physical background fields are non-zero, such as at
finite baryon density, we are led to a rich set of new interactions.
For example, we find interactions that mediate neutrino-photon
interactions in nuclear matter as discussed previously in
Ref.\cite{HHH2}.

A third role for these fields naturally suggests itself, namely to
promote them to the dynamical vector meson fields of QCD in the spirit
of vector meson dominance.  We then have an interesting set of pCS
terms which couple the vector mesons of QCD to fundamental gauge
fields.  A detailed analysis of the phenomenology of these terms will
be presented elsewhere~\cite{HHH1}.  We mention here 
that the couplings we find lead to results
for both the rate and polarization structure of the decay $f_1
\rightarrow \rho^0 \gamma$ which are in agreement with experiment.
We emphasize that the processes resulting from the identification
of the background fields with vector mesons
are not to be thought of as interactions in the fundamental,
underlying theory of quarks and leptons.  
Rather, they emerge in a low-energy effective description 
of QCD coupled to electroweak gauge fields.  

Since in the real world the $W$ and $Z$
bosons are much heavier than the scale of QCD, the $W$ and $Z$ should
be integrated out and replaced by the charged and neutral currents to
which they couple.  
In the formal limit $g_{1,2}\to 0$, the $W$ and $Z$ are explicit 
degrees of freedom in the low-energy theory, and the constraints of
$SU(2)_L\times U(1)_Y$ gauge symmetry are explicit.  
Since we remain in a perturbative regime, the constraints survive
at the physical values of these couplings.  
Restricting the weak currents to their components
involving light fields then gives a low-energy effective
Lagrangian which is valid for energies and momentum transfers 
up to the scale at which chiral perturbation theory breaks down, 
$4 \pi f_\pi \sim 1\, {\rm GeV}$.  

The centerpiece of our analysis 
is the derivation of a new counterterm which
must be added to the WZW term in order to maintain gauge
invariance in the presence of background fields.  
Once the counterterm is fixed, the full action  
provides a complete description of the global current anomalies 
of the theory, and also leads to a rich set of new interactions
with many potentially important physical applications. 

\vskip 0.2in
\noindent
{\bf Acknowledgments}
\vskip 0.1in
\noindent
We thank Jon Rosner and Koichi Yamawaki
for helpful discussions.
Research supported by the U.S.~Department of Energy  
grant DE-AC02-76CHO3000 and by NSF Grants PHY-00506630 and 0529954.



\begin{thebibliography}{99}  

\bibitem{Wess}  
  J.~Wess and B.~Zumino,
  ``Consequences of anomalous Ward identities,''
  Phys.\ Lett.\ B {\bf 37}, 95 (1971).

\bibitem{Witten}
  E.~Witten,
  ``Global Aspects Of Current Algebra,''
  Nucl.\ Phys.\ B {\bf 223}, 422 (1983).

\bibitem{steinberger} 
  J.~Steinberger,
  ``On the use of subtraction fields and the lifetimes of some types of meson
  decay,''
  Phys.\ Rev.\  {\bf 76}, 1180 (1949).
  
\bibitem{BJ} 
  J.~S.~Bell and R.~Jackiw,
  ``A PCAC puzzle: $\pi^0 \to \gamma \gamma$ in the sigma model,''
  Nuovo Cim.\ A {\bf 60}, 47 (1969).

\bibitem{Adler}
  S.~L.~Adler,
  ``Axial vector vertex in spinor electrodynamics,''
  Phys.\ Rev.\  {\bf 177}, 2426 (1969).

\bibitem{Bardeen}
  W.~A.~Bardeen,
  ``Anomalous Ward Identities In Spinor Field Theories,''
  Phys.\ Rev.\  {\bf 184}, 1848 (1969).


\bibitem{KRS}
  O.~Kaymakcalan, S.~Rajeev and J.~Schechter,
  ``Nonabelian Anomaly And Vector Meson Decays,''
  Phys.\ Rev.\ D {\bf 30}, 594 (1984).

\bibitem{Manohar:1984uq}
  A.~Manohar and G.~W.~Moore,
  ``Anomalous Inequivalence Of Phenomenological Theories,''
  Nucl.\ Phys.\  B {\bf 243}, 55 (1984).

\bibitem{Kawai:1984mx}
  H.~Kawai and S.~H.~H.~Tye,
  ``Chiral Anomalies, Effective Lagrangian And Differential Geometry,''
  Phys.\ Lett.\  B {\bf 140}, 403 (1984).

\bibitem{Hill2}
  C.~T.~Hill and R.~J.~Hill,
  ``T-parity violation by anomalies,''
  arXiv:0705.0697 [hep-ph], {\it Phys. Rev. D, in press};   
  C.~T.~Hill and R.~J.~Hill, 
  ``Topological physics of little higgs bosons,''
  Phys.\ Rev.\  D {\bf 75}, 115009 (2007).

\bibitem{GW}
  J.~Goldstone and F.~Wilczek,
  ``Fractional Quantum Numbers On Solitons,''
  Phys.\ Rev.\ Lett.\  {\bf 47}, 986 (1981).

\bibitem{Harada} See the review of
  M.~Harada and K.~Yamawaki,
  ``Hidden local symmetry at loop: A new perspective of composite gauge boson
  and chiral phase transition,''
  Phys.\ Rept.\  {\bf 381}, 1 (2003),
  and references therein.

\bibitem{HHH1}
J.~A.~Harvey, C.~T.~Hill and R.~J.~Hill,
work in progress.

\bibitem{HHH2}
J.~A.~Harvey, C.~T.~Hill and R.~J.~Hill,
``Anomaly mediated neutrino-photon interactions at finite baryon density,''
arXiv:0708.1281 [hep-ph], {\it Phys. Rev. Lett., in press}.

\bibitem{MiniBooNE}  A.~A.~Aguilar-Arevalo {\it et al.}  
[MiniBooNE Collaboration],
  Phys.\ Rev.\ Lett.\  {\bf 98}, 231801 (2007).

\bibitem{Kaiser}
  N.~Kaiser and U.~G.~Meissner,
  ``Generalized hidden symmetry for low-energy hadron physics,''
  Nucl.\ Phys.\  A {\bf 519}, 671 (1990).
 
 \bibitem{Klingl}
  F.~Klingl, N.~Kaiser and W.~Weise,
  ``Effective Lagrangian approach to vector mesons, their structure and
  decays,''
  Z.\ Phys.\  A {\bf 356}, 193 (1996).
  
 \bibitem{Truhlik}
  E.~Truhlik, J.~Smejkal and F.~C.~Khanna,
  ``Electromagnetic isoscalar $\rho$ $\pi$ $\gamma$ 
  exchange current and the  anomalous
  action,''
  Nucl.\ Phys.\  A {\bf 689}, 741 (2001).

\bibitem{dhokerI}
E.~D'Hoker and E.~Farhi,
``Decoupling A Fermion Whose Mass Is Generated By A Yukawa Coupling: The
General Case,''
Nucl.\ Phys.\  B {\bf 248}, 59 (1984).

\bibitem{dhokerII}
E.~D'Hoker and E.~Farhi,
``Decoupling A Fermion In The Standard Electroweak Theory,''
Nucl.\ Phys.\  B {\bf 248}, 77 (1984).

\bibitem{cth}
   C.~T.~Hill and C.~K.~Zachos,
  ``Dimensional deconstruction and Wess-Zumino-Witten terms,''
  Phys.\ Rev.\ D {\bf 71}, 046002 (2005);
  C.~T.~Hill,
  ``Exact equivalence of the D = 4 gauged Wess-Zumino-Witten term and the D = 5
  Yang-Mills Chern-Simons term,''
  Phys.\ Rev.\ D {\bf 73}, 126009 (2006);
   C.~T.~Hill,  
   ``Anomalies, Chern-Simons terms and chiral delocalization in extra
   dimensions,''
   Phys.\ Rev.\ D {\bf 73}, 085001 (2006).

\bibitem{Coleman:1969sm}
  S.~R.~Coleman, J.~Wess and B.~Zumino,
  ``Structure of phenomenological Lagrangians. 1,''
  Phys.\ Rev.\  {\bf 177}, 2239 (1969);
C.~G.~Callan, S.~R.~Coleman, J.~Wess and B.~Zumino,
  ``Structure of phenomenological Lagrangians. 2,''
  Phys.\ Rev.\  {\bf 177}, 2247 (1969).

\bibitem{Chu:1996fr}
  C.~S.~Chu, P.~M.~Ho and B.~Zumino,
  ``Non-Abelian Anomalies and Effective Actions for a Homogeneous Space
  $G/H$,''
  Nucl.\ Phys.\  B {\bf 475}, 484 (1996).

\bibitem{Witten2} 
  C.~G.~Callan and E.~Witten,
  ``Monopole Catalysis Of Skyrmion Decay,''
  Nucl.\ Phys.\  B {\bf 239}, 161 (1984).

\bibitem{hillc}
  C.~T.~Hill,
  ``Topological solitons from deconstructed extra dimensions,''
  Phys.\ Rev.\ Lett.\  {\bf 88}, 041601 (2002).

\bibitem{rjh}
  R.~J.~Hill, 
  ``$SU(3)/SU(2)$: the simplest WZW term,'' in preparation. 

\bibitem{Domokos}
  S.~K.~Domokos and J.~A.~Harvey,
  ``Baryon number-induced Chern-Simons couplings of vector and axial-vector
  mesons in holographic QCD,''
  Phys.\ Rev.\ Lett.\  {\bf 99}, 141602 (2007).

\end{thebibliography}
\end{document}